\documentclass[a4paper,twoside]{article}
\newcommand{\affil}[1]{$^{\rm #1}$}
\usepackage[authoryear]{natbib}
\bibpunct{(}{)}{;}{a}{}{,}
\usepackage{graphicx}
\date{} %Please leave the date blank
\newcommand{\Hi}{\textup{H\,{\mdseries\textsc{i}}}}
\newcommand{\HI}{\textup{H\,{\mdseries\textsc{i}}}~}

\newcommand{\Msun}{{M$_{\odot}$}}

\def\kms{{km~s$^{-1}$}}

\def\degr{\hbox{$^\circ$}}
\def\arcmin{\hbox{$^\prime$}}

\title{\large\bf\flushleft Is Centaurus A special? A neutral hydrogen perspective}
\author{\parbox{\textwidth}{\flushleft
\vspace{-0.5cm}
{\it Christian Struve\affil{A,B,D}, Raffaella Morganti\affil{A,B}, Tom A. Oosterloo\affil{A,B}, and Bjorn H.C. Emonts\affil{C}}\\
\vspace{0.4cm}
{\small \affil{A}\,Netherlands Foundation for Research in Astronomy, Postbus 2, 7990 AA Dwingeloo, The Netherlands}\\
{\small \affil{B}\,Kapteyn Astronomical Institute, University of Groningen, PO Box 800, 9700 AV Groningen, The Netherlands}\\
{\small \affil{C}\,CSIRO Australia Telescope National Facility, PO Box 76, Epping NSW, 1710, Australia}\\
{\small \affil{D}\,Email: struve@astron.nl}}}
\begin{document}
\twocolumn[
\begin{changemargin}{.8cm}{.5cm}
\begin{minipage}{.9\textwidth}
\vspace{-1cm}
\maketitle
%
%
%%%%%%%%%%%%%     ABSTRACT    %%%%%%%%%%%%%
%Abstract of no more than 200 words here.
\small{\bf Abstract:} Due to the proximity, the neutral hydrogen belonging to Centaurus~A can be observed at high resolution with good sensitivity. This  allows to study the morphology and kinematics in detail in order to understand the evolution of this radio-loud source (e.g. merger history, AGN activity). At the same time, it is important to compare the results to other sources of the same class (i.e. early-type galaxies in general and radio galaxies in particular) to see how Centaurus~A fits into the global picture of early-type/radio galaxy evolution.
The amount of \Hi , the morphology of a warped disk with \HI clouds surrounding the disk and the regular kinematics of the inner part of the \HI disk are not unusual for early-type galaxies. The growing evidence that mergers are not necessarily responsible for AGN activity fits with the observational result that the recent merger event in Centaurus~A is not connected to the current phase of activity. Based on these results, we conclude that Centaurus~A has typical neutral hydrogen properties for an early-type and radio galaxy and it can therefore --- from an \HI perspective --- be seen as a typical example of its class.

%%%%%%%%%%%%%     KEYWORDS    %%%%%%%%%%%%%
\medskip{\bf Keywords:} galaxies: active --- galaxies: ellipticals --- galaxies: individual (Centaurus A) --- galaxies: kinematics and dynamics --- galaxies: structure --- galaxies: ISM
% Please write all keywords in lower case. PASA uses the
% standard list of subject headings adopted by The Astrophysical Journal
% and available from http://www.journals.uchicago.edu/ApJ/keywords_text.html.
% Keywords are separated by em-dashes, i.e. ---

%%%%%%%%DO NOT EDIT%%%%%%%%%%%%
\medskip
\medskip
\end{minipage}
\end{changemargin}
]
\small
%%%%%%%%EDIT FROM HERE%%%%%%%%%%%%

% CAN WE MOVE FIGURE 1 & 2 FROM THE END TO SOMEWHERE IN THE MIDDLE???????

\begin{figure*}[h]
\begin{center}
\includegraphics[scale=0.45]{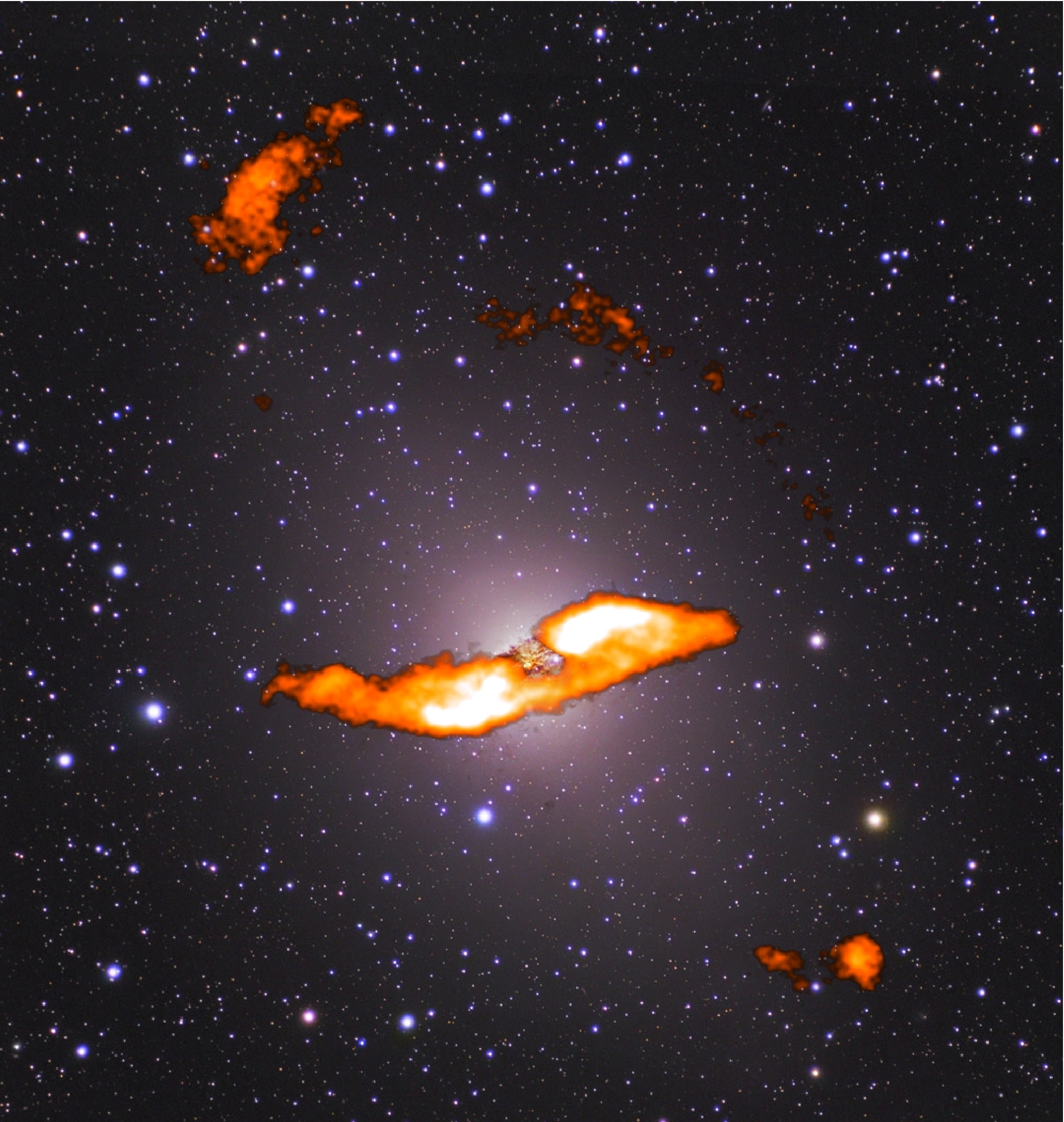}
\caption{Large-scale \HI emission (in orange) from the ATCA observations (C. Struve et al. in prep.) overlaid on an optical image (Credit: Robert Gendler \& Stephane Guisard). The newly discovered, unresolved cloud (detected in emission) is located to the NE of the disk.}
\label{dss.m0}
\end{center}
\end{figure*}

\begin{figure*}[h]
\begin{center}
\includegraphics[scale=0.6, angle=270]{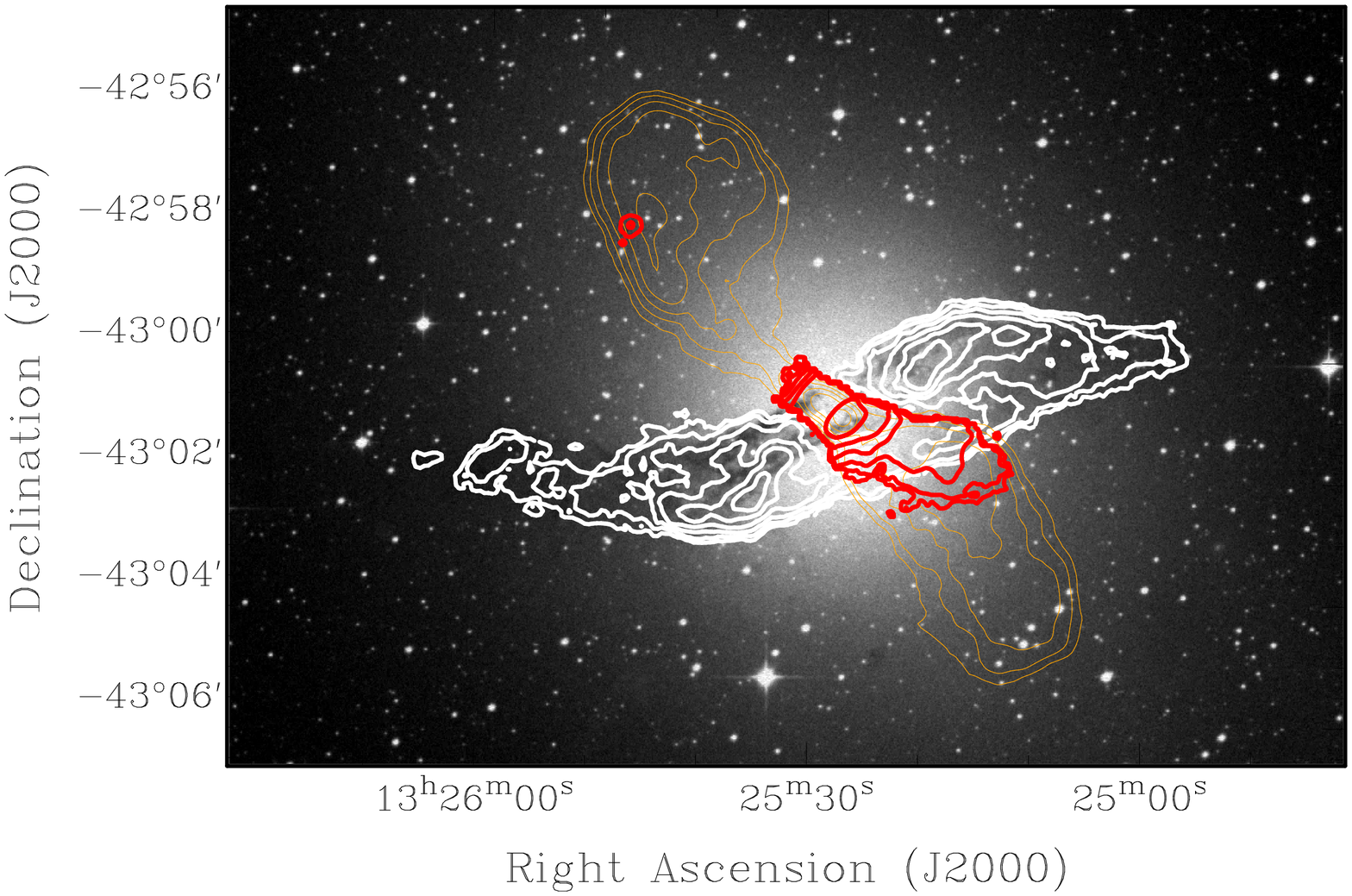}
\caption{Zoom-in of Fig.~\ref{dss.m0}. DSS-red image of overlaid with the \HI emission (white contours) and absorption (red contours) and the inner radio lobe (orange contours). The figure also shows the newly discovered cloud detected in absorption against the northern lobe.}
\label{fig2}
\end{center}
\end{figure*}

\section{Introduction}

The role of the gas in the formation and evolution process of early-type galaxies is still not fully understood. For example, recent studies have shown a large complexity in the gas structures in these systems (e.g. Morganti et al. 2006, Sarzi et al. 2006, Combes et al 2007) despite their often unspectacular optical appearance. In addition, some sources show nuclear activity while others do not. Cold-gas structures represent a fossil record of the formation and evolution of early-type galaxies. In particular, gas found on kiloparsec scales can be used to trace the evolution of the host galaxy (e.g. major merger vs. small accretions). In that respect, neutral hydrogen is an important tracer of these events as it often extends the dust and optical (disk) structure by a factor of two or more.

Close to the centre of galaxies ($<100$~pc), (cold) gas also plays a crucial role as it can provide the fuel that is needed to make the central black hole active. It is often believed that mergers are important in driving gas to the centre (see e.g. Hibbard \& van Gorkom 1996, Barnes 2002), but recent studies of radio galaxies have shown that the activity in some galaxies may be associated with the (slow) accretion of gas from the ISM/IGM (e.g. Best et al. 2005).

Despite the need for large statistical investigations to understand (and classify) the different mechanisms at work (e.g. mergers, interactions, accretion, AGN activity etc.), it is indispensable to study close-by objects in great detail with very high linear resolution. For example, the circumnuclear region around the black hole ($< 100$~pc) can only be resolved in the most nearby galaxies to a degree that is needed to understand the accretion/fueling process.

By far the closest radio-loud early-type galaxy is Centaurus~A (NGC~5128) at a distance of only 3.8~Mpc\footnote{At this distance 1\arcmin ~corresponds to $\sim 1.1$~kpc.} (Harris 2009). Cen~A has been studied in all possible wavelength regimes with very high linear resolution (for an overview see the review by Israel (1998) and the other contributions to this volume). However, it is crucial to compare Cen~A with other sources to check whether it is a typical example of its class, or whether it is unusual w.r.t. other radio-loud, low-luminosity sources.

In this paper we discuss whether Cen~A is special as seen from the neutral hydrogen perspective. That is, do the Cen~A properties (such as e.g. \HI mass, morphology, kinematics etc.) differ from other early-type galaxies and does Cen~A share properties with radio galaxies that have comparable luminosity? In Sect.~2 we give a brief description of the \HI morphology and kinematics on kpc and sub-kpc scales. Section~3 compares Cen~A with other nearby early-type galaxies and in Sect.~4 Cen~A is compared to a complete sample of nearby radio galaxies. We summarize our comparison in Sect.~5.

\section{Cen~A as seen in neutral hydrogen}

\subsection{The \HI large-scale morphology and kinematics}

Van Gorkom et al. (1990) performed the first interferometric observations to map the \HI in emission and absorption, they showed that the \HI follows the prominent, warped dust lane (e.g. Nicholson et al. 1992, Quillen et al. 2006) and that most of the \HI rotates around the centre. Some hints of unsettled gas were found in the outer SE parts of the disk suggesting that some of the gas has not yet settled into regular rotating orbits.

\HI well outside the disk was discovered by Schiminovich et al. (1994). A number of \HI clouds located 10 to 15~kpc from the nucleus form a partial ring structure with a smooth N-S velocity gradient, rotating perpendicular to the gas in the dust lane, but in the same sense as the main stellar body of the galaxy. The prominent (warped) dust and gas disk in the central region, the outer optical shell structure, together with the partial \HI ring structure indicate that Cen~A is likely the product of a recent accretion of a small gas-rich spiral galaxy with a larger elliptical galaxy as already suggested by Baade \& Minkowski 1954) and Tubbs (1980).

We have recently performed (with the ATCA) new, higher resolution observations with better sensitivity to study the circumnuclear region and disk kinematics in more detail (C. Struve et al. in prep.). The new observations have allowed to better separate emission and absorption and to detect new faint features. Figure~\ref{dss.m0} shows the large-scale \HI emission of Centaurus~A, Fig.~\ref{fig2} is a zoom-in and shows the emission and absorption of the disk. Our main results are:
\begin{itemize}
   \item \HI is detected in emission along the dust lane with a diameter of 13~kpc. The total \HI mass in emission is M$_{\HI}=4.6\cdot 10^8$~\Msun . Absorption is detected against the nucleus, the northern jet and against the southern lobe.
   \item Tilted-ring modeling shows that the inner, modestly warped ($<30$\degr ) 5~kpc disk in radius can be well described by a set of tilted circular rings that explain the morphology and kinematics. The rotation curve quickly rises in the central part and remains essentially flat thereafter with only a very modest decline with radius. At larger radii asymmetries in the morphology and kinematics are present and the gas has not yet settled into regular rotating orbits.
   \item Our tilted-ring model also describes the dust disk detected by Spitzer Quillen et al (2006) and the stellar ring recently discovered by Kainulainen et al. (2009).
   \item We detect no emission/absorption down to about 3.6~mJy~beam$^{-1}$ in the inner 1~kpc (except for the central beam) in agreement with observations at other frequencies (Nicholson et al. 1992, Quillen et al. 2006).
   \item There is no need for non-circular motions for $1<r<5$~kpc.
   \item Based on the regular rotation for $r<5$~kpc, we estimate the age of the disk to be a few times $10^8$~yr and hence the event that produced the disk is too old to explain the current phase of activity ($<10^7$~yr). However, the disk could have approximately the same age as estimated for the northern middle lobe (Saxton et al. 2001).
   \item A significant fraction of the absorption detected against the core (see below) cannot be explained by the rotating large-scale disk and hence this absorbing gas must be located close to the nucleus.
   \item Outside the disk we discovered two additional clouds that could be part of the partial ring structure discovered by (Schiminovich et al. 1994).
\end{itemize}

\subsection{The circumnuclear region}

Dedicated \HI absorption studies of the nucleus previously have shown that the absorption against the nucleus is solely redshifted (e.g. van der Hulst et al. 1983, Sarma et al. 2002). This was taken as evidence that the absorbing \HI might fall towards the black hole, potentially providing the fuel that is needed for the nuclear activity. However, our new observations (C. Struve et al. in prep.; Morganti et al. 2008) show the existence of blueshifted absorption and that the redshifted absorption is significantly broader ($\Delta v_{\rm{absorp}}^{\rm{total}} = 400$~\kms ) than previously measured. The broad absorption component was previously missed due to insufficient bandwidths. Morganti et al. (2008) suggest that the nuclear \HI absorption could be interpreted as a circumnuclear disk which would be the atomic counterpart of the observed CO disk (Liszt 2001, Espada et al. 2009).

Jones et al. (1996) have shown that the core at VLBI scales is visible at 8.4~GHz, but at lower frequencies the core becomes self-absorbed. Hence, the central \HI absorption detected with the ATCA occurs against an extended structure. New Long Baseline Array observations show a similar velocity width (compared to the ATCA observations) of the deepest part of the absorption ($\Delta v_{\rm{absorp}}^{\rm{deep}} \approx 70$~\kms ) against the bright part of the beginning VLBI jet (projected distance $\sim 1$~pc from the core). However, this absorbing gas could be part of the kpc-scale disk as it is close in velocity to the systemic velocity of Cen~A. A further analysis is needed to clarify the physical origin of this absorption.

\section{Comparison with other early-type galaxies}

Although early-type galaxies used to be perceived to be gas poor, different gas phases are in fact detected in many objects, provided deep observations are available. Ionised gas has recently been found in $75$\% in early-type galaxies (Sarzi et al. 2006) and molecular gas was detected in up to 54\% of the observed sources (e.g. Combes et al. 2007, Flaquer et al. 2008). The presence of ionised, molecular and atomic gas in Cen~A is therfore not surprising (for a summary of the gas properties in Cen~A see Morganti 2009).

Also in \HI more than 50\% of non-cluster galaxies are detected with \HI masses between $10^6$ and $10^{10}$~\Msun ~(Morganti et al. 2006, Oosterloo et al. 2007). A variety of gas morphologies is present, ranging from very extended (up to 200~kpc) regularly rotating disk/ring structures of low column density (Oosterloo et al. 2007) to long tidal tails and barely resolved blobs ($<4$~kpc diameter) (Morganti et al. 2006). Disks in early-type galaxies can form simultaneously with their host, from major mergers of gas-rich galaxies (see e.g. Hibbard \& van Gorkom 1996, Barnes 2002). In those mergers, about half the gas is quickly funneled to the centre, partly consumed in a burst of star formation and partly settling in a nuclear central disk (Bournaud et al. 2005). The other half is ejected to large distances but might remain bound to the merger remnant and eventually will fall back settling in a large-scale disk. Compared to other galaxies, the presence of a large-scale rotating \HI disk with unsettled gas in the outer parts of the disk (at $r>5$~kpc), as well as the partial ring structure at larger distances from the nucleus is not unusual.

The observed, essentially flat and only mildly decreasing \HI rotation curve is a commonly observed phenomenon in other early-type (disk) galaxies (see e.g. Noordermeer et al. 2007). In addition, some early-type galaxies (e.g. NGC~3108, ESO~381-47, IC~2006) have ring structures, i.e. they have a depression of \HI towards the nucleus (Oosterloo et al. 2002, Donovan et al. 2009, Franx et al. 1994). Therefore, neither the rotation curve nor the radial \HI surface density distribution of Cen~A is unusual when compared with other early-type galaxies.

One of the well known characteristic of the gas/dust disk in Cen~A is the warped structure. While the origin of warps (e.g. mergers, interactions, accretion of gas from the intergalactic medium) is still under debate and may differ from object to object (for a discussion see e.g. Briggs 1990, Garc{\'{\i}}a-Ruiz et al. 2002), it is an observational fact that most disk galaxies are at least mildly warped in \HI (Garc{\'{\i}}a-Ruiz et al. 2002). Most extended \HI disks/rings in early-type galaxies are warped, e.g. in IC~4200 or ESO~381-47 (Serra et al. 2006, Donovan et al. 2009). In some cases, the warping amplitude can be large (i.e. approaching 90\degr ) and also start well within the optical disk as is observed in NGC~2685 (J{\'o}zsa et al. 2009). Considering the merger history of Cen~A, it is therefore not unusual that the gas disk in Cen~A is warped (Sect.~2), nor the warping amplitude is spectacularly high. It is rather its proximity and the orientation on the sky that make Cen~A appear as a peculiar warped object.

\begin{figure}[h]
\begin{center}
\includegraphics[scale=0.45, angle=270]{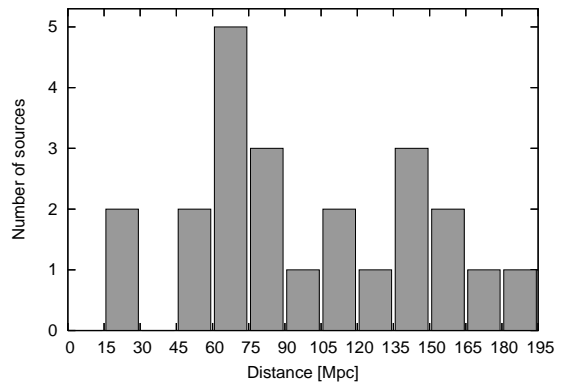}
\caption{Distances of the radio sources of the complete sample of nearby radio galaxies (Emonts 2006, B.H.C. Emonts et al. in prep.). Cen~A ($D=3.8$~Mpc) is located a factor $>4$ closer than the closest sample source and a factor $\sim 26$ than the average sample source ($\langle D_{\rm{sample}} \rangle =101$~Mpc).}
\label{distances}
\end{center}
\end{figure}
\begin{figure}[h]
\begin{center}
\includegraphics[scale=1.2]{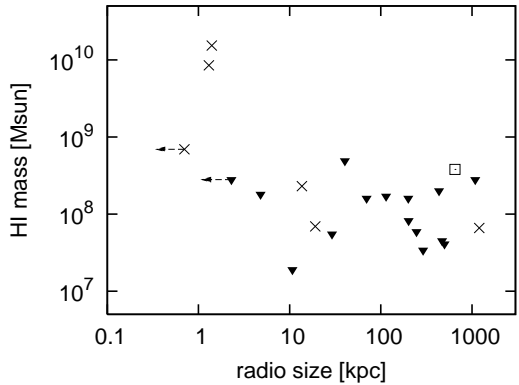}
\caption{Total \HI mass detected in emission plotted against the linear size of the radio sources of the complete sample of radio galaxies (B.H.C. Emonts et al. in prep.). The crosses denote the \HI detections, the filled triangles the upper limits of the non-detections. The open square represents Cen~A.}
\label{segregation}
\end{center}
\end{figure}

\section{Comparison with other radio galaxies}

In the previous section we have shown that Cen~A has an \HI morphology and kinematics that is also found in a number of other early-type galaxies and that in this respect Cen~A is not special. However, it is still important to compare Cen~A with other radio sources as it could be an exceptional radio galaxy.

Radio activity is a short lived (sometimes recurrent) phenomenon (typically believed to be $<10^8$~yr) in the evolution of an early-type galaxy. It is therefore natural that the number density of radio galaxies in the (local) universe is significantly lower than that of early-type galaxies in general. This makes Cen~A a unique object as it can be observed in great detail because it is much closer to us than any other radio source with comparable radio luminosity. Figure~\ref{distances} shows the distances of a complete sample of nearby low-luminosity radio sources (Emonts 2006, B.H.C. Emonts et al. in prep.). The distance advantage of Cen~A (and hence the linear resolution) is evident from the figure as even the closest sample source is located more than 4 times as far away as Cen~A. The sample covers all nearby ($z<0.41$) sources of a large fraction of the northern sky with $F_{\rm{408MHz}}>0.2$~Jy with the restriction that cluster members are excluded as the probability of an \HI detection in those environments is low. For details of the sample and the analysis we refer the reader to (Emonts 2006, Emonts et al. 2007).

Six out of the 21 sample sources ($=29$\%) are detected in emission. The reason for the slightly lower detection rate (compared to radio-quiet early-type galaxies, $>50$\%) is that the upper \HI mass limits of the non-detections are higher. Therefore, the difference is believed to be the result of observational limitations and not to be an intrinsic feature of radio galaxies (Emonts 2006, B.H.C. Emonts et al. in prep.). Similar to the sample of early-type galaxies, a variety of \HI structures is found for the radio galaxies, including (settling) disk structures. However, the striking result from this analysis is that large amounts of \Hi, rotating in regular disks, are only detected around {\sl compact} sources (typically $<10$~kpc, Emonts et al. 2007). Extended radio sources --- comparable to Cen~A --- do only show modest amounts of \HI (few times $10^8$~\Msun ), Fig.~\ref{segregation} (see also Emonts et al. 2007). If Cen~A would be located at a distance of $70$~Mpc the peak of the emission would correspond to about $4\sigma_{\rm{rms}}$ resulting in a blob like structure, extended over 1-2 beams. In addition, the deep part of the absorption would be detected. However, at larger distances ($D > 70$~Mpc), the emission of Cen~A would not have been detected. Therefore, the results from the statistical sample are not in conflict with Cen~A's \HI mass.

Cen~A, with its 650-kpc continuum structure (see e.g. Feain 2009), is believed to have gone through multiple phases of AGN activity. Recurrent radio activity is also found in a number of other sources (see e.g. Schoenmakers et al. 2000). We note that at least one \HI detected compact radio sample source shows a 250~kpc relic structure at the sub-mJy level (B2~0258+35; C. Struve et al. in prep.), showing that also some of these compact sources had previous phases of AGN activity. Therefore, also the radio continuum structure of Cen~A does not appear unusual compared to other radio galaxies.

\HI in absorption has been detected in a large number of radio galaxies (e.g. Vermeulen et al. 2003, Morganti et al. 2005). Initially, only absorption profiles redshifted (relative to the systemic velocity) were found and the absorbing gas clouds were seen as evidence for infall towards the nuclear region, potentially providing the fuel for the AGN (van Gorkom et al. 1989). This picture has changed in recent years with the availability of more sensitive and broader-band observations revealing also blueshifted absorption (Vermeulen et al. 2003, Morganti et al. 2005). In some cases, the \HI absorption is centred on the systemic velocity of the galaxy and is often interpreted as a circumnuclear disk/torus, (see e.g. Conway \& Blanco 1995, van Langevelde et al. 2000, Peck \& Taylor 2001) which is actually in agreement with theoretical predictions (e.g. Maloney et al. 1996). The \HI ATCA observations of Cen~A are in agreement with a circumnuclear structure (see Morganti et al. 2008) which is further supported by the existence of a molecular circumnuclear disk (Liszt 2001, Neumayer et al. 2007).

\section{Summary}

In order to understand Centaurus~A in the context of galaxy formation and evolution, we have compared the \HI properties of Cen~A with early-type and radio galaxies. The \HI mass, its distribution and the mainly settled kinematics is commonly found in other early-type/radio galaxies. The current phase of AGN activity is not connected to the recent merger which is in line with recent results for a sample of radio galaxies. The absorption against the nucleus is red- and blueshifted with respect to the systemic velocity and is in agreement --- as is also the case in other sources --- with a circumnuclear \HI disk/torus structure. Hence, Centaurus~A seems to be --- from an \HI perspective --- a typical galaxy of its class.

\section*{Acknowledgments}

This research was supported by the EU Framework 6 Marie Curie Early Stage Training programme under contract number MEST-CT-2005-19669 ``ESTRELA''.

%\end{multicols}

\end{document}